\documentclass[doublecol]{epl2}
\usepackage{amsmath}
\usepackage{amsfonts}
\usepackage{amssymb}
\usepackage{graphicx}
\usepackage{dcolumn}
\usepackage{bm}

\title{Sudden jumps  and  plateaus in the quench dynamics of a Bloch state }

\author{J.~M.~Zhang\inst{1,2} \and Hua-Tong Yang\inst{3} }

\institute{
  \inst{1} Fujian Provincial Key Laboratory of Quantum Manipulation and New Energy Materials,
College of Physics and Energy, Fujian Normal University, Fuzhou 350007, China\\
  \inst{2} Fujian Provincial Collaborative Innovation Center for Optoelectronic Semiconductors and Efficient Devices, Xiamen, 361005, China \\
  \inst{3} School of Physics, Northeast Normal University, Changchun 130024, China
}

\abstract{We take a one-dimensional tight binding chain with periodic boundary condition and put a particle in an arbitrary Bloch state, then quench it  by suddenly  changing the potential of an arbitrary site. In the ensuing time evolution, the probability density of the wave function at an arbitrary site \emph{jumps indefinitely between plateaus}. This phenomenon adds to a former one in which the survival probability of the particle in the initial Bloch state shows \emph{cusps} periodically, which was found in the same scenario [Zhang J. M. and Yang H.-T., EPL, \textbf{114} (2016)  60001]. The plateaus support the scattering wave picture of the quench dynamics of the Bloch state. Underlying the cusps and jumps is  the exactly solvable, nonanalytic dynamics of a Luttinger-like model, based on which, the locations of the jumps and the heights of the plateaus are accurately predicted.}


\begin{document}
\maketitle

\section{Introduction}

Singularities can show up in the time evolution of a physical quantity. A classic example is provided by a model solved rigorously by Stey and Gibberd four decades ago \cite{stey}. The model is in the Lee-Friedrichs class \cite{friedrichs,lee}, and is actually a paradigm for quantum decay \cite{cohen,fermi,scienceopen}. It consists of an equidistant quasi-continuum extending from $-\infty $ to $ + \infty $, and an extra discrete level, which couples to all the continuum levels with the same strength. Let us initialize the system on the discrete level and let it decay into the quasi-continuum. It turns out that the survival probability of the initial state shows cusps periodically in time. More recently, in the surge of nonequilibrium dynamics of many-body systems \cite{bloch02,rigol,weiss,eisert,nonequilibrium}, Heyl \emph{et al.} have discovered non-analyticities at critical times in the time evolution of some Loschmidt echo related quantities \cite{heyl}. In contrast to the former example, the non-analyticities were sought deliberately by noting the formal similarity between the Loschmidt amplitude $\langle \Psi_i |e^{-i Ht }| \Psi_i \rangle$ and the partition function Tr$(e^{-\beta H})$, and by borrowing notions from equilibrium phase transitions. The singularities were then legitimately called dynamical quantum phase transitions. Thereafter, many systems demonstrating this novel type of phase transition in the time domain were found \cite{dutta, huang, schuricht, heyl14, heyl15, dora}.

We have also recently found some nonsmooth dynamics in a very simple scenario \cite{bloch}. The setting is a one-dimensional tight binding chain with periodic boundary condition, which is arguably the simplest model in solid state physics \cite{mermin}. Initially a particle is put in some Bloch state with an arbitrary momentum. Then suddenly the potential of some site is changed. In the subsequent evolution, both the survival probability and the reflection probability, which correspond to the particle remaining in its initial state and being momentum reversed, respectively, show cusps periodically.  The cusps were explained by exactly solving an idealized model. The mechanism  is completely different from those functioning in the examples above.

The singularities invite experimental verifications. Unfortunately, so far this goal is yet to be fulfilled. For the model of Stey and Gibberd, a clean realization is to put a two-level atom in a multi-mode cavity and to position it at an appropriate point \cite{parker, meystre, ligare95, ligare02}. However, to the best of our knowledge, this has never been attempted experimentally, possibly due to the challenge of the necessity of precisely locating the atom. An alternative scheme is to use the one-dimensional tight binding model \cite{scienceopen}. But the drawback is that it is confined to the perturbative regime. As for the dynamical quantum phase transition, the essential difficulty is that the central quantity, namely the Loschmidt amplitude for a many-body system, cannot be measured directly. Therefore, the phase transition has to be inferred from some conventional quantities which are experimentally accessible. Some proposals in this spirit do exist \cite{heyl14, heyl15, huang} but experimental realization is still missing.

Motivated by this problem, we have reinvestigated the singular dynamics of the Bloch state \cite{bloch}.
From the experimental point of view, to verify the singular dynamics by measuring directly the two probabilities (or populations) is not necessarily an easy task. In many circumstances, it is much easier to measure \emph{local} quantities. We have thus investigated the time evolution of the probability density of the wave function at an arbitrary site. From the theoretical point of view, this study is also worthy as it provides a real space  picture of the dynamics, complementary to the momentum space picture established previously. It turns out that the finding is as interesting as in the momentum space. On a large time scale, the trajectory of the local density is characterized by plateaus and sudden switches between them. The durations of the plateaus depend on the site. In particular, the heights of the plateaus are very sensitive to the location of the site; they vary significantly even between adjacent sites. Therefore, we believe measuring the local density is a much more feasible means to verify the singular dynamics of the particle.

\section{Sudden jumps and plateaus}\label{sec2}

\begin{figure*}[tb]
\centering
\includegraphics[width= 0.45 \textwidth ]{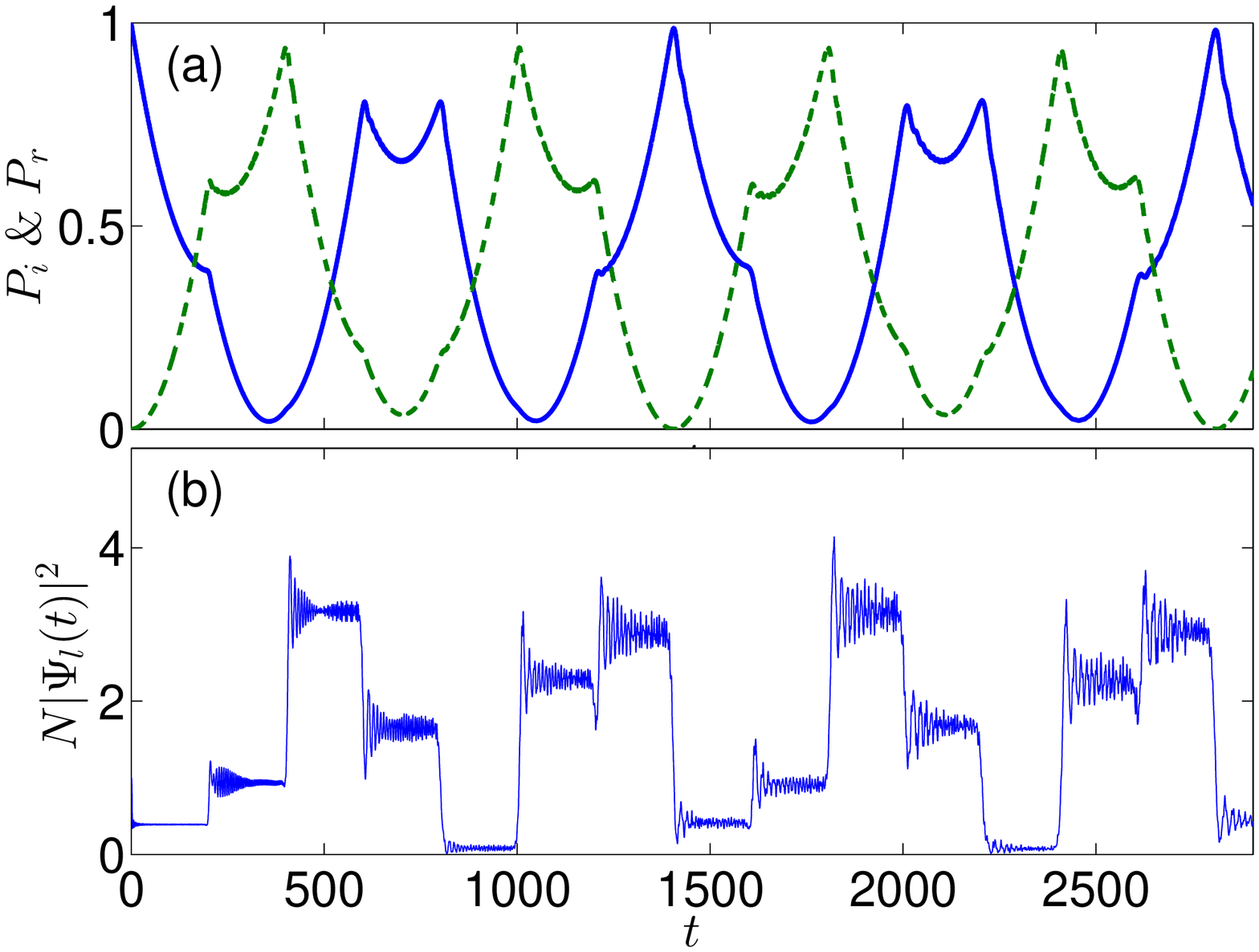}
\hspace{0.25cm}
\includegraphics[width= 0.45 \textwidth ]{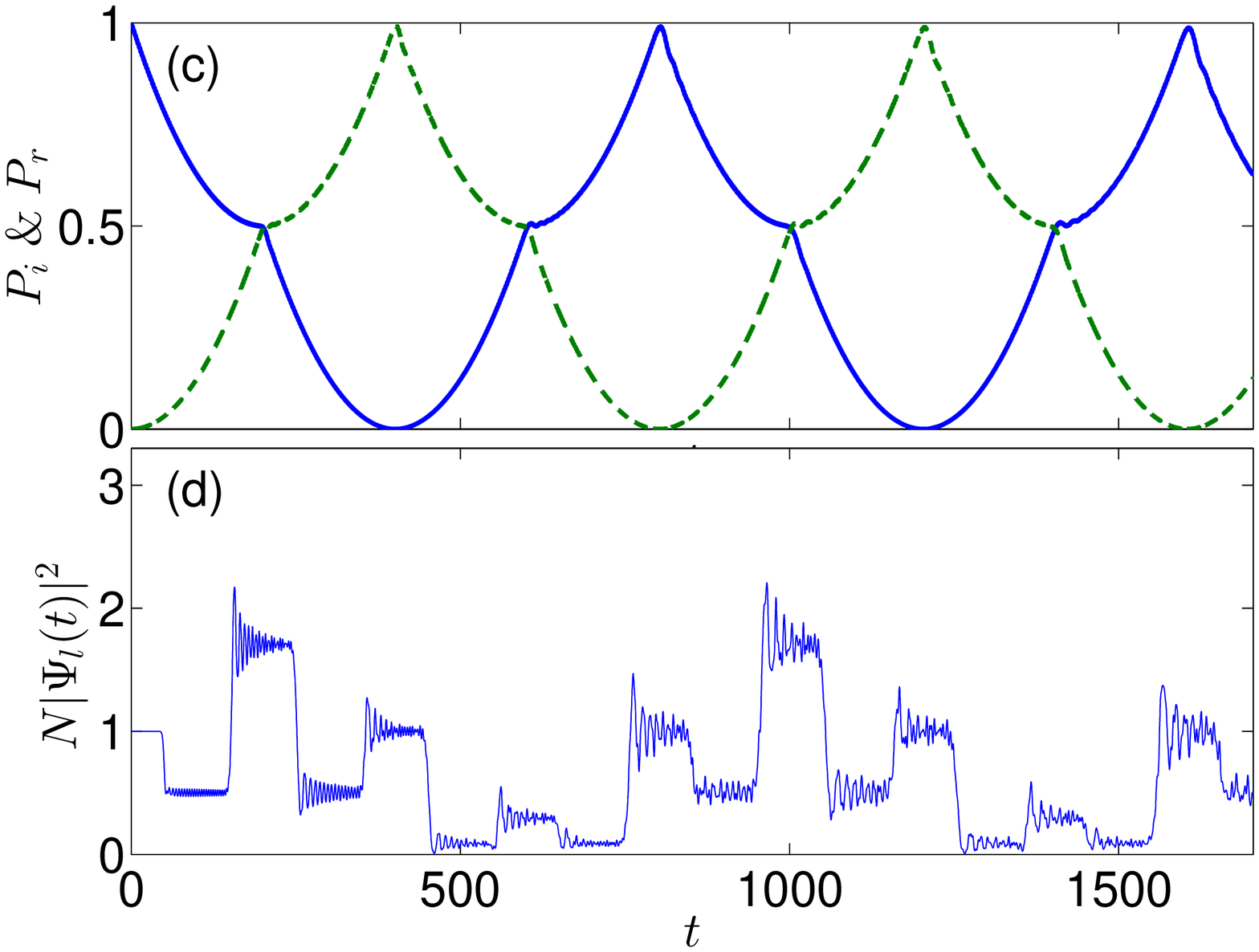}
\caption{(Color online) Time evolution of the survival probability $P_i$ (solid lines), the reflection probability $P_r$ (dashed lines), and the local probability density $|\Psi_l |^2 $ [see Eqs. (\ref{pipr}) and (\ref{probden}) for definition]. The left two panels correspond to one case, while the right two panels to another case. The common parameters are $N = 401$, $k_i = 100$, while the remaining parameters are  $(U, l ) = (2.5, 5 )$ in left two panels, and $(U, l) = (2, 100)$ in the right two panels.
\label{mom_real}}
\end{figure*}

The setting is a one-dimensional tight binding chain with  periodic boundary condition. The original Hamiltonian is ($\hbar = 1$ throughout this paper, and the hopping strength between two adjacent sites is set as the unit of energy)
\begin{eqnarray}\label{h0}
  \hat{H}_0  &=& - \sum_{l=0}^{N-1 } (|l\rangle \langle l+1 | + |l +1 \rangle \langle l | ) ,
\end{eqnarray}
where $|l\rangle $ denotes the Wannier function at site $l $. By the periodic boundary condition, $|l+ N \rangle = | l\rangle $.
The so-called Bloch states are simultaneously eigenstates of $\hat{H}_0 $ and eigenstates of the translation operator $\hat{T} |l\rangle = | l+1 \rangle $. They are indexed by the integer $k$, and have the explicit expression of
\begin{eqnarray}\label{bstate}
  \langle l | k \rangle  &=& \frac{1}{\sqrt{N }} e^{i q l } ,
\end{eqnarray}
in the real space. Here $q = 2 \pi k /N $ is the so-called wave vector. Apparently, $|k \rangle = | k + N \rangle $.
The corresponding eigenenergies are $\varepsilon(q) = -2 \cos q $. Like the Wannier functions $|l\rangle $ are the basis vectors in the real space, the Bloch states $|k\rangle $ are the basis vectors in the momentum space.

Now the scenario is as follows. Initially a particle is put in some arbitrary Bloch state $|k_i \rangle $, i.e., $|\Psi (0)\rangle = |k_i \rangle $. Then at $t= 0$, the potential of some site is changed to $U $ suddenly and held fixed afterwards. Without loss of generality, the site is assumed to be the  $l=0 $ one. The final Hamiltonian which will govern the evolution of the wave function $|\Psi(t) \rangle $ of the particle is then
$ \hat{H}_f  = \hat{H}_0   + \hat{H}_1 $, with $ \hat{H}_1 =   U |0\rangle \langle 0 |$. The characteristic feature of the newly introduced perturbation is that it couples two arbitrary Bloch states with the same strength, i.e.,
\begin{eqnarray}\label{g}
 g =  U/N  = \langle k_1 | \hat{H}_1 | k_2 \rangle
\end{eqnarray}
regardless of the values of $k_{1,2}$, as can be easily checked by using the explicit expression of (\ref{bstate}).

The question is how does the wave function $|\Psi(t) \rangle $ evolve. An intuitive picture is that the particle will be reflected back and forth between the two groups of Bloch states centered at $|+ k_i\rangle $ and $|-k_i \rangle $.
Therefore, in a previous work \cite{bloch}, the two quantities
\begin{eqnarray}\label{pipr}
  P_i (t) = |\langle +k_i | \Psi(t) \rangle |^2, \quad     P_r (t) = |\langle -k_i | \Psi(t) \rangle |^2,
\end{eqnarray}
which are called the survival probability and the reflection probability, respectively, were studied. The finding is that,
both $P_i$ and $P_r $ show cusps periodically. Examples are shown in figs.~\ref{mom_real}(a) and \ref{mom_real}(c). In ref.~\cite{bloch}, the cusps and even the whole trajectories of $P_{i,r}$ were explained by identifying and solving an ideal model
behind them.

Quantities like $P_{i,r}$ are defined in terms of the momentum space. They are motivated by the Rabi oscillation picture above. From another point of view, the newly introduced barrier at site $l= 0$ will excite scattering waves, which will propagate both forwards and backwards away from the barrier. Because the group velocity of a wave packet on the lattice chain is upper-bounded, it will take finite times for the scattering waves to reach a site. Once the scattering waves come back to the barrier, they will generate secondary scattering waves, and so on. Therefore, the wave function $ |\Psi(t) \rangle  $ might exhibit rich temporal and spatial structures.

We have thus investigated numerically the time evolution of the probability density
\begin{equation}\label{probden}
 D_l (t ) \equiv |\langle l | \Psi (t) \rangle |^2 = | \Psi_l (t) |^2
\end{equation}
of the wave function at an arbitrary site $l$. Corresponding to figs.~\ref{mom_real}(a) and \ref{mom_real}(c), we have figs.~\ref{mom_real}(b) and \ref{mom_real}(d), respectively, where the plateaus and the fast switch between them are
the most prominent features. They are not rigorously plateaus---there are fluctuations. But in each interval, the curve fluctuates apparently around a horizontal line. In figs.~\ref{mom_real}(a) and \ref{mom_real}(b), where $l= 5 \ll N = 401 $, it seems that the sudden jumps occur simultaneously with the cusps. However, this is not the case in figs.~\ref{mom_real}(c) and \ref{mom_real}(d), where $l / N \simeq 1/4 $.  Further numerical examples indicate that the times when the jumps occur depend on the location of the site under observation. This is in line with the scattering wave picture above.

It should be stressed that the temporal patterns, i.e., the cusps and the plateau switches, are very robust. Although in fig.~\ref{mom_real}, we have shown only the first few cusps and jumps in a limited time interval ensuing the sudden quench, actually they persist \emph{forever}. Moreover, their sharpness never deteriorates with time, which strongly hints at the regularity of the dynamics of the model in question.

\section{Explanation}\label{explanation}

\begin{figure}[tb]
\centering
\includegraphics[width= 0.45 \textwidth ]{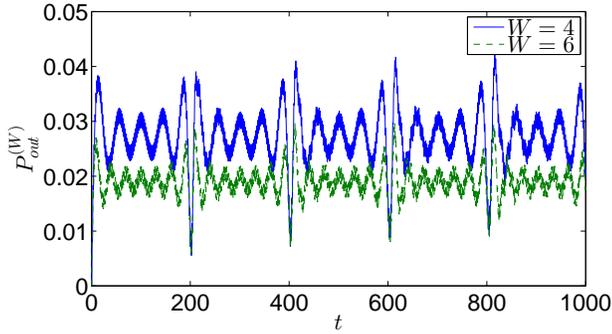}
\caption{(Color online) Time evolution of the population on the Bloch states outside the two slots of $|k-k_i| \leq W$ and $|k+k_i | \leq W $ [see eq.~(\ref{pout}) for definition]. The parameters are the same as in the left two panels of fig.~\ref{mom_real}, i.e., $N = 401 $, $k_i = 100$, and $ U = 2.5 $.
\label{pop}}
\end{figure}

\begin{figure}[tb]
\centering
\includegraphics[width= 0.45 \textwidth ]{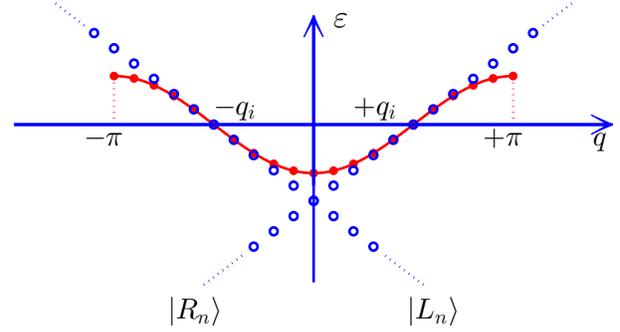}
\caption{(Color online) The red solid curve depicts the dispersion relation $\varepsilon (q) = -2 \cos q $ of the unperturbed Hamiltonian (\ref{h0}). The red solid dots on the curve indicate the Bloch states (\ref{bstate}). The blue hollow dots, which form two linear branches, indicate the levels in the ideal model. Note that while for the realistic model (\ref{h0}), the number of levels is finite and the value of the wave vector $q$ is confined to the interval $[-\pi, +\pi]$, in the ideal model, the number of levels is infinite and $q$ extends from $-\infty $ to $+\infty $.
\label{luttinger}}
\end{figure}


Behind the plateaus and sudden jumps is actually the same ideal model which is responsible for the cusps \cite{bloch}. This ideal model was motivated and justified by an important numerical observation. That is, for a wide range of parameters, only those Bloch states with energies in the vicinity of the energy of the initial Bloch state $|k_i \rangle $ participate significantly in the dynamics. This means two groups of Bloch states, with wave vectors either in the vicinity of $+ q_i$ or $-q_i$. Here $q_i = 2\pi k_i/ N$ is the wave vector of the initial Bloch state. Figure \ref{pop} demonstrates this point very well. There, for a lattice of size $N = 401 $, and an initial Bloch state with $k_i = 100$, the population on those Bloch states outside of the two slots of $|k -k_i| \leq W $ and $ |k+k_i | \leq W $ is studied. That is, the quantity
\begin{eqnarray}\label{pout}
  P_{out}^{(W)}(t) \equiv  \sum_{ |k -k_i| > W \atop |k+k_i |>W  } | \langle k | \Psi(t)\rangle   |^2
\end{eqnarray}
is followed. Here $W$ is the cutoff. It is seen that even for $W = 4$, the two narrow slots, which cover less than $3\%$ of the Brillouin zone, almost exhaust the probability.

Therefore, it is tempting to construct a fictitious model, which resembles the realistic model $\hat{H}_f$ in the two slots but can deviate from it (even significantly) outside. Around the two points of $\pm q_i$, the dispersion relation of the unperturbed Hamiltonian (\ref{h0}) can be linearized. The wave vectors are equally spaced with a gap $\delta = 2\pi/ N $ and the eigenenergies are equally spaced with a gap $\Delta = \varepsilon'(q_i ) \delta =  (4 \pi \sin q_i  )/ N $. Hence, we truncate the realistic spectrum of $\hat{H}_0$ by retaining only the two slots and then extend them into two branches of linear spectra $\{ |R_n\rangle \}$ and $\{ |L_n\rangle \}$, $-\infty < n < + \infty $. See fig.~\ref{luttinger}. For $n$ in the vicinity of zero, these states correspond to the Bloch states as
\begin{eqnarray}\label{correspondence}
  |R_n \rangle  \leftrightarrow  |k_i + n \rangle , \quad |L_n \rangle \leftrightarrow   |- k_i - n \rangle .
\end{eqnarray}
For $|n|$ sufficiently large, this correspondence fails and the states $|R_n \rangle $ and $|L_n \rangle $ are indeed fictitious.
Like the Bloch states $|k \rangle $ are eigenstates of $\hat{H}_0 $,  the basis states $\{ |R_n\rangle \}$ and $\{ |L_n\rangle \}$  are eigenstates of the fictitious unperturbed Hamiltonian $\hat{\mathcal{H}}_0 $,
\begin{eqnarray}\label{fh0}
  \hat{\mathcal{H}}_0 |R_n \rangle = n \Delta |R_n \rangle, \quad  \hat{\mathcal{H}}_0 |L_n \rangle = n \Delta |L_n \rangle.
\end{eqnarray}
One should note that exactly the same energy spectra occur in the Luttinger model \cite{luttingerbook}. Of course, here it is just a single-particle problem. Resembling (\ref{g}), the fictitious perturbation $ \hat{\mathcal{H}}_1 $ couples two arbitrary states with the same strength
\begin{eqnarray}\label{fh1}
  g  = \langle R_{n_1}| \hat{\mathcal{H}}_1| R_{n_2} \rangle = \langle L_{n_1}| \hat{\mathcal{H}}_1| L_{n_2} \rangle = \langle R_{n_1}| \hat{\mathcal{H}}_1| L_{n_2} \rangle.
\end{eqnarray}
Equations (\ref{fh0}) and (\ref{fh1}) define the fictitious model completely. The idea of its construction is that, the exact details of the original Hamiltonian $\hat{H}_f  $ beyond the two slots do not matter much, as the population on those levels is negligible. Therefore, to some extent, it is free to rectify that part to make analytic calculation possible. This will be justified if the prediction of this fictitious model agrees with the exact results.

It should be stressed that although in the original problem, there are only a finite number of ($N$ actually) levels, here in the fictitious model, we have two branches of levels, each consisting of an \emph{infinite} number of levels. Moreover, in the original model, Bloch states $|k\rangle $ and $|k + N \rangle $ are identified as the same state, while here
\begin{subequations}
\begin{eqnarray}
   \langle R_{n_1} |L_{n_2} \rangle =0 ,  \;\;\; \quad \quad \quad \quad 
   \quad   & & \forall \; n_1 , n_2 ,  \\
 \langle  R_{n_1} |R_{n_2}\rangle =\langle  L_{n_1} |L_{n_2}\rangle = 0, \; & & \text{ if } n_1 \neq n_2 .
\end{eqnarray}
\end{subequations}
Because eventually we are to calculate the amplitude of the wave function at an arbitrary site $l$, we have to postulate the expressions of the basis states $\{ |R_n\rangle \}$ and $\{ |L_n\rangle \}$ in the real space. This is done simply by generalizing (\ref{correspondence}) to all $n$'s, and taking (\ref{bstate}), i.e.,
\begin{eqnarray}\label{realspace}
  \langle l | R_n \rangle  &=&  \frac{1}{\sqrt{N}} \exp[ i (q_i + n \delta ) l ] =  \langle l | L_n \rangle^* .
\end{eqnarray}

In ref.~\cite{bloch}, the dynamics of the initial state
$  |\Psi(0)\rangle  =  |k_i \rangle = | R_0 \rangle $ under the control of the fictitious Hamiltonian $ \hat{\mathcal{H}}_0 +\hat{\mathcal{H}}_1 $
has been solved. In terms of the states with definite parities,
$  |A_n^{\pm} \rangle =  (|R_n \rangle \pm  |L_n\rangle )/\sqrt{2 } $,
the state at time $t$ is in the form of
\begin{equation}\label{form}
  |\Psi(t)\rangle = \frac{1}{\sqrt{2}} |A_0^- \rangle + \frac{1}{\sqrt{2}} \sum_{n=-\infty}^\infty  \psi_n (t ) |A_n^+ \rangle.
\end{equation}
At this point we have to introduce a very important quantity, i.e., the so-called Heisenberg time $T \equiv 2\pi /\Delta $. This is the time which sets the time scale of the dynamics in question. It is actually the period between the cusps in figs.~\ref{mom_real}(a) and \ref{mom_real}(c). For $r T < t < (r+ 1) T$, where $r$ is a nonnegative integer, the value of $\psi_0 $ is ($s  \equiv  t - r T $)
\begin{equation}\label{psi0}
  {\psi}_{0}(t) =  \left( 1 - \frac{i 2 g s  }{1+ i g T } \right)   e^{-i  r \theta } ,
\end{equation}
and the value of $\psi_n $ ($n\neq 0 $) is
\begin{equation}\label{psin}
  {\psi}_{n}(t) =   \frac{2g}{(1+ i g T)\Delta   } \left( \frac{e^{-i n \Delta s } - 1}{ n }  \right) e^{-i  r \theta } .
\end{equation}
Here $e^{-i \theta } = (1- i gT )/( 1+ i g T )$ is a phase factor. In (\ref{psi0}), we see that $\psi_0 $ is a continuous but nonsmooth function of time $t$. Its trajectory on the complex plane is like the trajectory of a ball inside a circular billiard. This explains the cusps in figs.~\ref{mom_real}(a) and \ref{mom_real}(c), as $P_{i, r} = |1 \pm \psi_0 |^2/4$.
Substituting (\ref{realspace}), (\ref{psi0}), and (\ref{psin}) into (\ref{form}), we calculate straightforwardly
\begin{widetext}
\begin{eqnarray}\label{final}
 \sqrt{N}  \langle l | \Psi (t) \rangle &=& i \sin q_i l +  e^{-i  r \theta } \bigg[ \cos q_i l + \frac{g}{(1+ i g T)\Delta}\sum_{n=1}^\infty  \frac{e^{-i n \Delta s} - 1}{ n }  \left( e^{i q_i l + i n \delta l } + e^{-i q_i l - i n \delta l } \right) \bigg ] \nonumber  \\
   &=& i \sin q_i l +  e^{-i  r \theta } \bigg[  \cos q_i l + \frac{i g}{(1+ i g T)\Delta} \bigg( e^{i q_i l  } \big(  \beta (\delta l - \Delta s ) - \beta (\delta l)- \Delta s  \big) \nonumber \\
   & & \quad \quad \quad \quad\quad \quad \quad\quad \quad \quad \quad \quad\quad \quad \quad  +  e^{-i q_i l  } \big( \beta (-\delta l - \Delta s) - \beta (- \delta l)-\Delta s  \big ) \bigg) \bigg] \nonumber \\
   &=& i \sin q_i l +  e^{-i  r \theta } \bigg[  \cos q_i l + \frac{i g }{(1+ i g T)\Delta} \left( e^{i q_i l  } \gamma_+(s )
    +  e^{-i q_i l  } \gamma_-(s ) \right) \bigg] .
\end{eqnarray}
\end{widetext}

\begin{figure*}[tb]
\centering
\includegraphics[width= 0.4 \textwidth ]{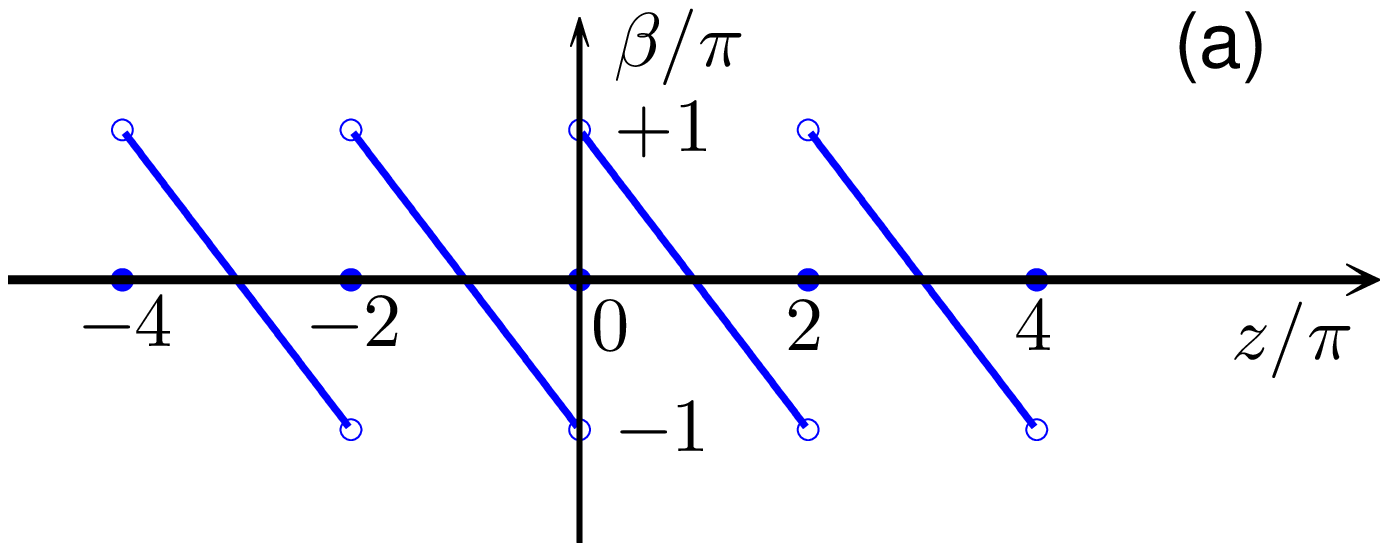}
\hspace{2cm}
\includegraphics[width= 0.4 \textwidth ]{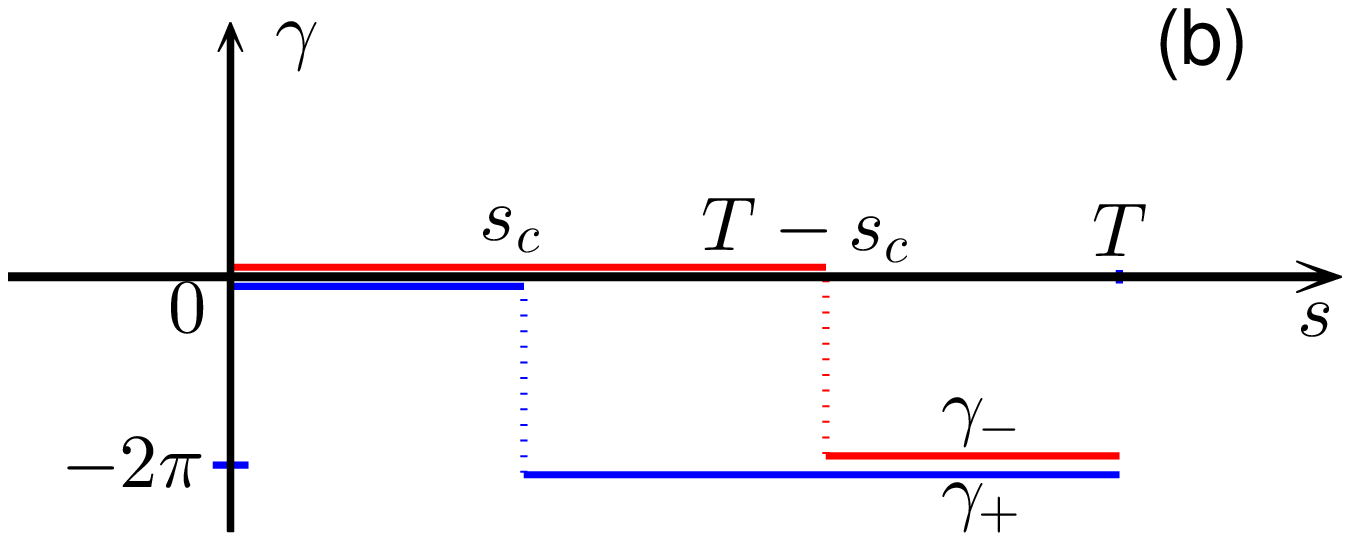}
\caption{(Color online) (a) Graph of the function $\beta(z)$, which is periodic and piecewise linear. (b) Graphs of the functions $\gamma_\pm (s)$, which are defined on the interval $[0, T ]$ and piecewise constant. See Eqs.~(\ref{betadef})-(\ref{gammavalue}) for their definitions.
\label{betagamma}}
\end{figure*}

In the second line of (\ref{final}) we have defined the function ($ z\in \mathbb{R} $)
\begin{eqnarray}\label{betadef}
  \beta (z) &=& 2 \sum_{n=1}^\infty \frac{\sin n z }{n } .
\end{eqnarray}
It is apparently periodic with a period of $2 \pi $. In ref.~\cite{courant}, it is calculated to be [see fig.~\ref{betagamma}(a) for its graph]
\begin{eqnarray}\label{betavalue}
  \beta (z) &=& \begin{cases}   -(z - 2 \pi \lfloor z/2\pi \rfloor -\pi ), &  z \neq 2 \pi r  , \\ 0, &  z = 2\pi r  , \end{cases}
\end{eqnarray}
where $\lfloor \cdot \rfloor$ is the floor function. In the third line, we have defined the functions
\begin{subequations}\label{gammadef}
\begin{eqnarray}
  \gamma_+(s )  &\equiv & \beta (+\delta l - \Delta s ) - \beta (+\delta l)- \Delta s, \\
  \gamma_-(s ) & \equiv & \beta (-\delta l - \Delta s) - \beta (- \delta l)-\Delta s,
\end{eqnarray}
\end{subequations}
on the interval $[0, T]$. Using (\ref{betavalue}), it is easy to find that they are step functions [see fig.~\ref{betagamma}(b) for their graphs]. Specifically,
\begin{subequations}\label{gammavalue}
\begin{eqnarray}
  \gamma_+(s ) &=& \begin{cases}   0, &  0 < s <  s_c   , \\  -2 \pi, &  s_c < s < T  , \end{cases} \\
  \gamma_-(s )&=& \begin{cases}   0, &  0 < s <  T- s_c    , \\  -2 \pi, & T - s_c < s < T  , \end{cases}
\end{eqnarray}
\end{subequations}
where $s_c = \delta l /\Delta $. The physical meaning of $s_c$ is the time needed for a wave packet, whose wave vector is centered at $q_i$, to go from the barrier to the site $l$ in the forward direction. The reason is simply that the group velocity of the wave packet is exactly $ v =\varepsilon'(q_i) =  \Delta / \delta $. Similarly, the Heisenberg time $T =2 \pi
/\Delta = N/v $ is just the time needed for the wave packet to complete the loop and return to the barrier. The time $T -s_c$ in (\ref{gammavalue}b) is then simply the time needed for a wave packet with wave vector $-q_i$ to go from the barrier to the site $l$.

\begin{figure*}[tb]
\centering
\includegraphics[width= 0.45 \textwidth ]{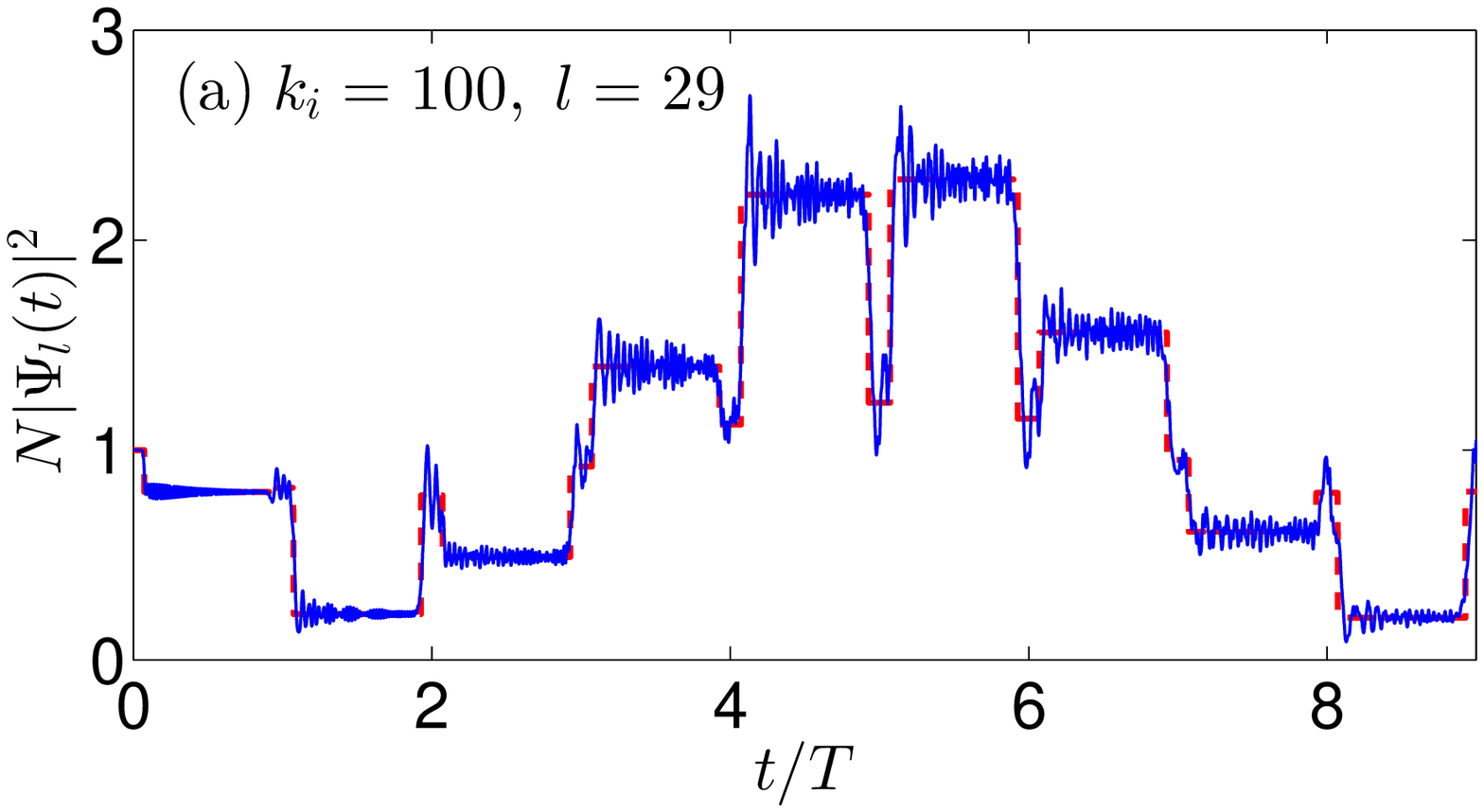}
\includegraphics[width= 0.45 \textwidth ]{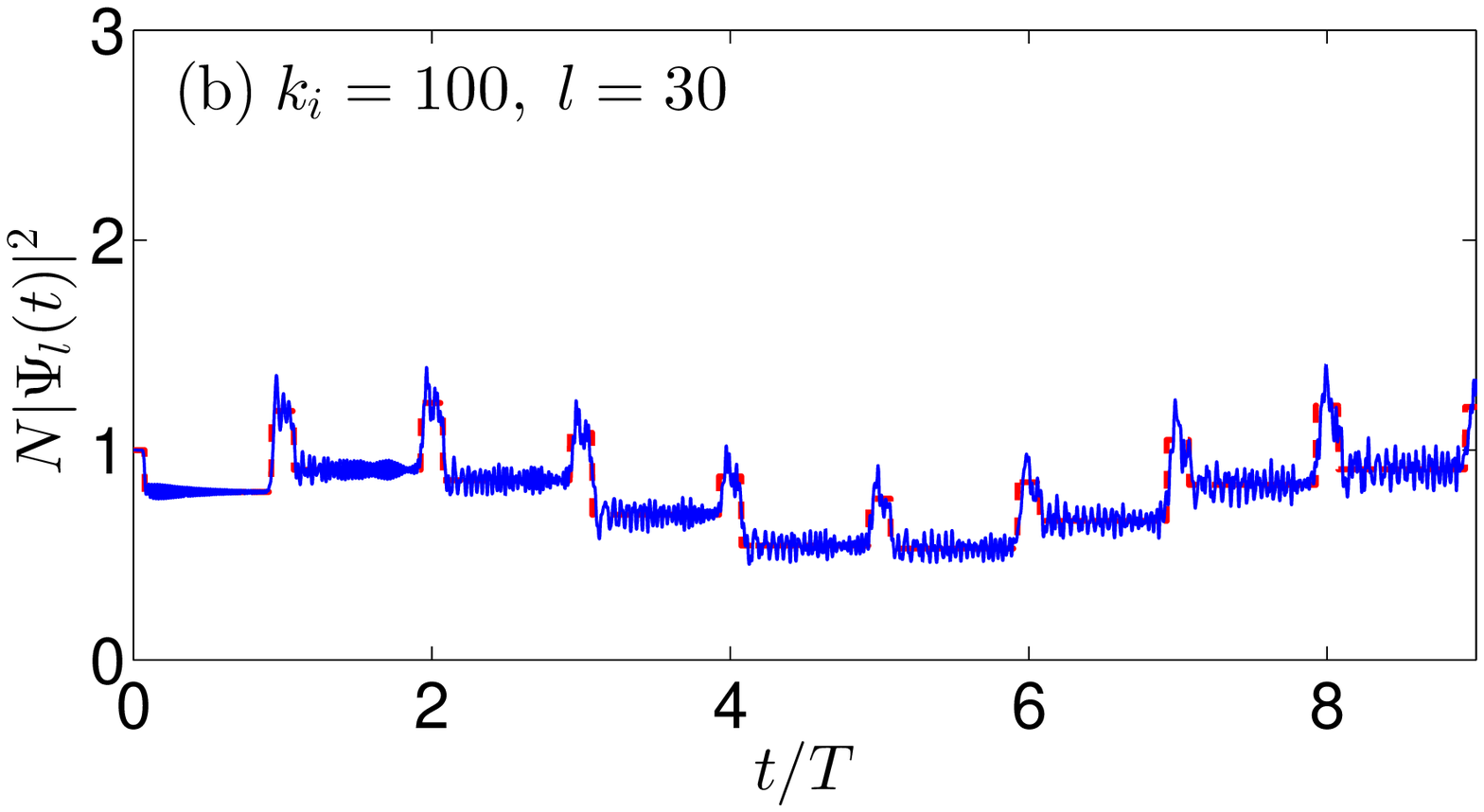}
\includegraphics[width= 0.45 \textwidth ]{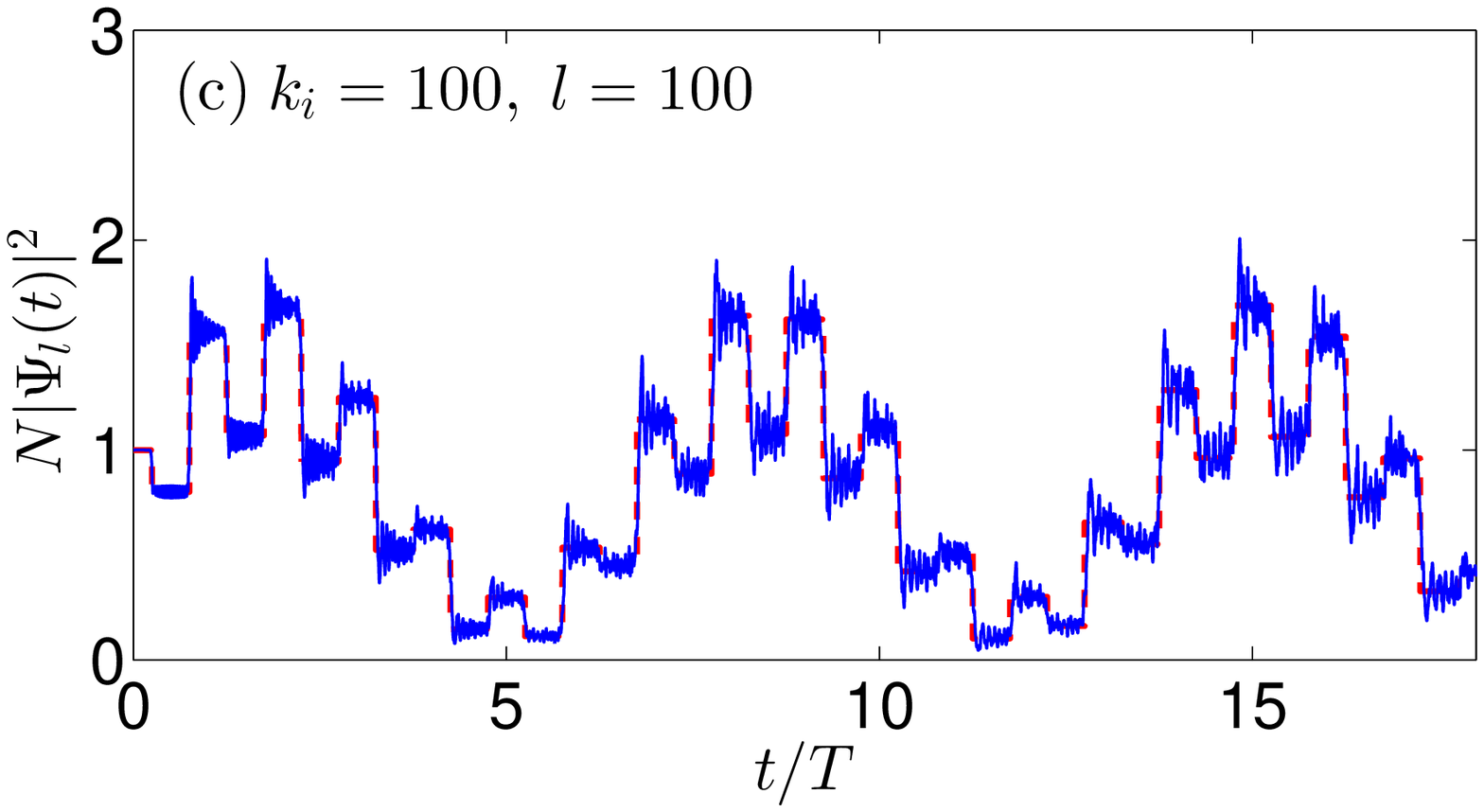}
\includegraphics[width= 0.45 \textwidth ]{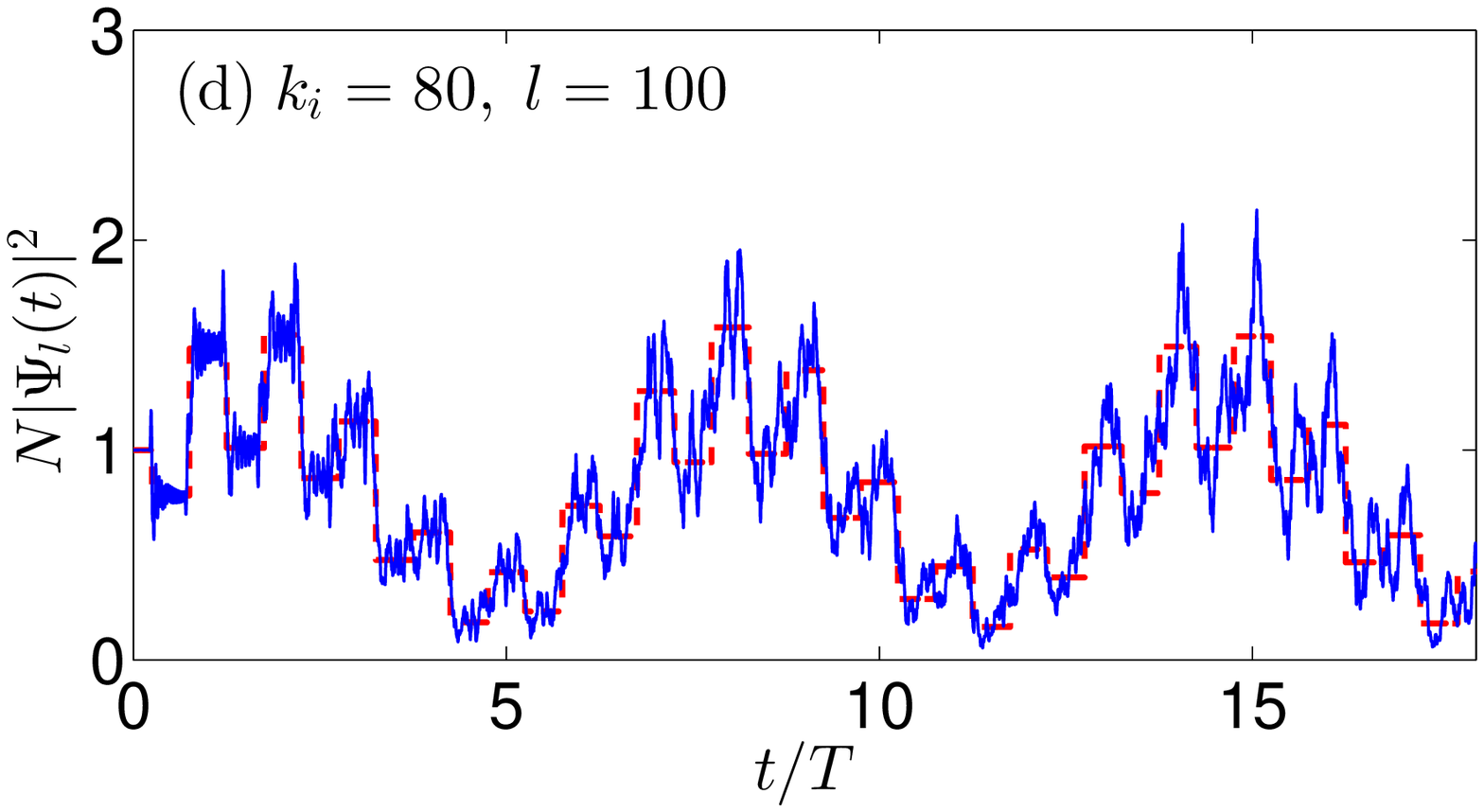}
\caption{(Color online) Comparison between the analytic predictions (red dashed lines) by (\ref{final}) and the numerical exact results (blue solid lines) for the local probability density. The common parameters are $N = 401$ and $U = 1$. The remaining parameters are shown in each panel. Panels (a) and (b) demonstrate that even between two adjacent sites, the evolution trajectories of the local probability densities can be very different. Panel (d) shows that even for large times, when the plateaus are not well shaped, the analytic predictions still capture the trend of evolution of $|\Psi_l |^2$ very well. Also note how the location of the observed site, namely the value of $l$, influences the lengths of the plateaus by comparing the upper panels with the lower ones.
\label{compare}}
\end{figure*}

Equation (\ref{final}) is our central result. We see that the amplitude $ \Psi_l (t)  $  jumps at $rT \pm s_c$. Between the jumps, the amplitude is a constant. Now we can compare the predictions of (\ref{final}), which are based on the fictitious model $\hat{\mathcal{H}}_0 + \hat{\mathcal{H}}_1$, and the numerical exact results, which are based on the realistic model $\hat{H}_0 +\hat{H}_1 $. This is done in fig.~\ref{compare}. We see that the analytic predictions based on the fictitious model (red dashed lines) agree with the
numerical results (blue solid lines) very well. Generally, not only the times when the jumps occur, but also the heights of the plateaus, are accurately predicted by (\ref{final}). An important prediction of (\ref{final}) is that, the value of $|\Psi_l(t)|^2 $ is very sensitive to the site index $l$, and this is vividly demonstrated in figs.~\ref{compare}(a) and \ref{compare}(b).

We should note the regularity of the time evolution of $|\Psi_l |^2$. The expression (\ref{final}) contains two parts. The first part, $i \sin q_i l $, is time independent. The second part is time dependent, but it factorizes into a part containing solely $r$ and a part containing solely $s$. Hence, $ \Psi_l (rT + s)  $  as a function of $r$ is of the form $a + b e^{-i r \theta }$, where $a$ and $b $ are constants. Its modulus squared is then a sinusoidal function of $r $. This harmonic oscillation feature is easily perceived in figs.~\ref{mom_real} and \ref{compare}. In particular, when $\theta / \pi $ is a rational number, as in fig.~\ref{mom_real}(d), where $\theta = \pi /2$, $ |\Psi_l (rT + s)|^2  $ is a periodic function of $r  $.

\begin{figure}[tb]
\centering
\includegraphics[width= 0.42 \textwidth ]{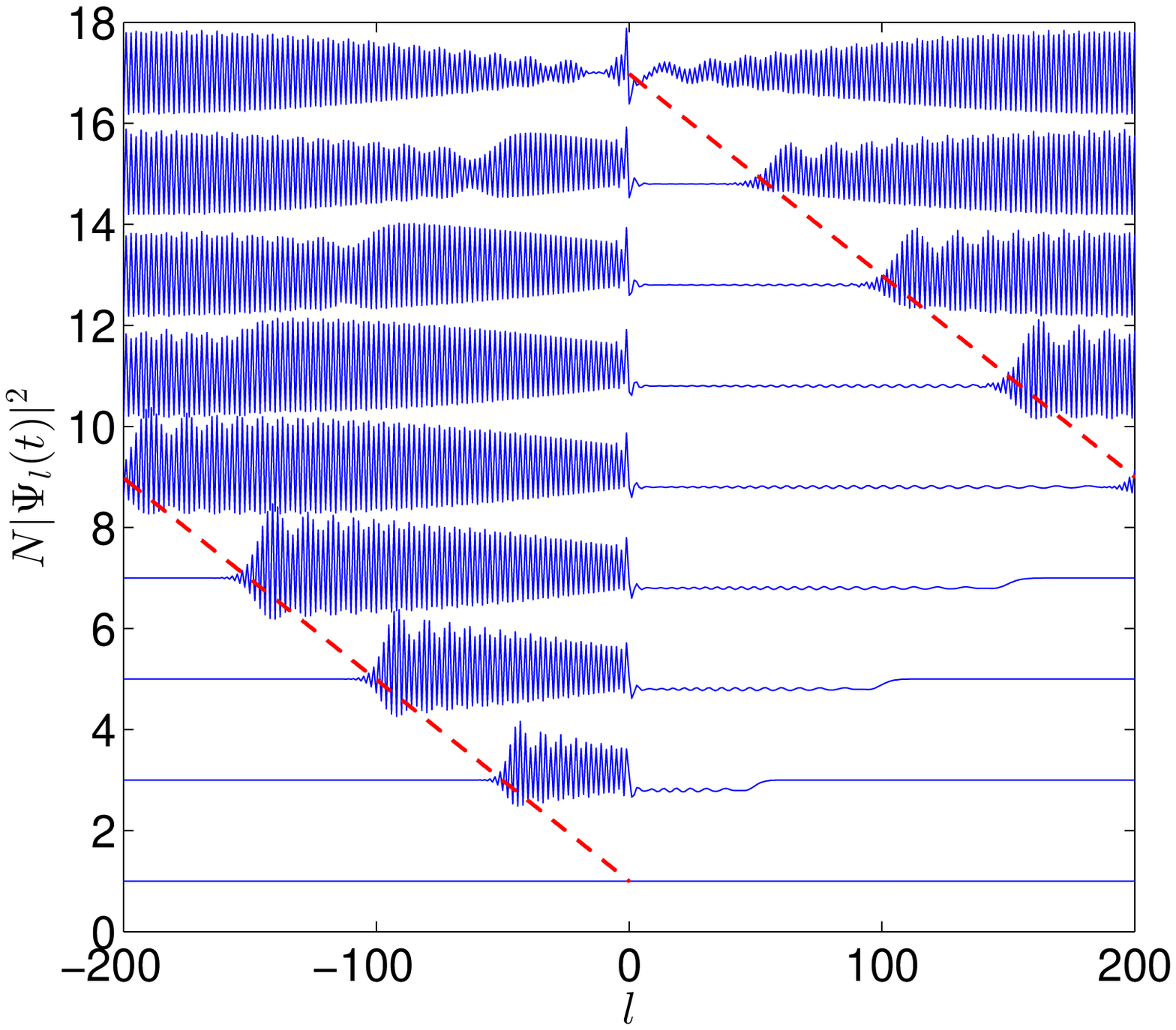}
\caption{(Color online) Nine equidistant snapshots of the probability distribution of the wave function in the time interval $[0, T]$. For clarity, from bottom to top, the $j$th curve is up shifted by $2(j-1)$. The parameters are $N = 401 $, $k_i =100$, and $U = 1$. The wave vector of the initial Bloch state is then $q_i = 2\pi k_i / N \simeq \pi/2$. The dashed line indicates the motion of the back scattering wave front with a constant velocity of $v= 2 \sin q_i \simeq  2 $. The forward scattering wave front is also clearly visible.
\label{snapshots}}
\end{figure}

The times when the jumps occur are in accord with the scattering picture below. Figure \ref{snapshots} shows the snapshots of the probability distribution of the wave function in the first period $[0, T]$. In the five snapshots at the bottom, the two wave fronts are readily recognized. They originate from the barrier and move at a constant velocity backwards and forwards. As the guiding dashed line shows, the velocity is exactly
\begin{eqnarray}
  v  &=& \frac{\partial \varepsilon}{\partial q}\big|_{q= q_i } = \frac{\Delta}{\delta } = 2\sin q_i ,
\end{eqnarray}
namely, the group velocity of a wave packet centered at $+ q_i $ or $- q_i $ in the momentum space.
Hence, the suddenly erected barrier generates forward scattering waves with wave vectors $q\simeq +q_i$, and back scattering waves with wave vectors $q\simeq - q_i $. The sudden jumps of $|\Psi_l |^2$ are then associated with the passing-by of their wave fronts of the site $l$. As can be seen in the lower few snapshots in fig.~\ref{snapshots}, in the forward direction, for a fixed site $l $, before the wave front reaches it, the value of $|\Psi_l |^2 $ is almost stationary at $1/N$; when the wave front crosses it, the value of $|\Psi_l |^2  $ experiences a swift change.

It takes the scattering waves exactly the time
\begin{eqnarray}
  \frac{N}{v}  &=& \frac{N}{\Delta/ \delta } = \frac{2 \pi }{\Delta } = T
\end{eqnarray}
to return to the barrier. Then the scattering waves will generate secondary scattering waves which also travel at the velocities $\pm v$ along the lattice. The times when the scattering waves, or the secondary ones, pass by the site $l$, are just $rT \pm s_c$, the times when the jumps occur.

\section{Experimental realization}\label{realization}

Like the van Hove singularity in the density of states of a solid \cite{vhs, li, solid}, here the singularity in the dynamics is also of interest by itself and  worth experimental verifications.

A promising approach to observing the sudden jumps and plateaus is to use coupled optical waveguides \cite{longhi}. Because of the quantum-optical correspondence, many coherent quantum phenomena have been successfully simulated using carefully engineered photonic guiding structures. In particular, some interesting phenomena with the tight binding model as setting, such as the dynamical localization \cite{dl, longhidl} and the Bloch oscillation \cite{pertsch, chiodo, morandotti}, have been realized.

In our case, the periodic boundary condition can be implemented by arranging the $N$ waveguides in a circular loop, and the initial Bloch state can be realized by carefully engineering the phases of the input laser beams, while the defect potential can be achieved by changing the refractive index of one of the waveguides, which can be done in several different ways \cite{longhi}.
The primary potential difficulty might come from the large time scale (the Heisenberg time $T $) involved, which is linearly proportional to the lattice size $ N $. In contrast, the time scales of the dynamical localization  and  the Bloch oscillation are independent of $N $.

\section{Conclusions and discussions}\label{conclusion}

We have reinvestigated the single-site quench dynamics of a Bloch state in a one-dimensional tight binding lattice, which was previously studied by us in ref.~\cite{bloch}. The motivation was on the one hand to get a complementary picture of the dynamics from the real space perspective, and on the other hand to find a quantity which is more convenient for measurement.

We have thus tracked the evolution of the probability density of the wave function at an arbitrary site, which is probably the most readily measurable quantity. It turns out to jump constantly, from plateau to plateau. In other words, its time development is not monotonic, but well structured. The times when the jumps occur, the durations as well as the heights of the plateaus, can all be accurately predicted by a fictitious model. Due to the nonlinearity of the realistic spectrum, and the finiteness of the number of levels in the realistic model, there are details beyond the predictions of the fictitious model, but the overall trend is well guided by its analytic predictions. The dependence of the times of the jumps on the site under consideration supports the scattering wave picture. Namely, the probability density jumps each time when a scattering wave passes by.

From the point of view of thermalization or equilibration, the model in question never thermalizes or equilibrates. First, in the figures, we can see that each time after the jump, the amplitude of fluctuation shrinks with time---it is equilibrating. However, this process is only to be interrupted by the next jump. The processes of jump and equilibration then alternate. Second, it is observed that even between two neighboring sites, the probability densities have completely different trajectories and never merge together. This means that the wave function will never become ``even'' across the lattice (like the water in a tank). Hence, the dynamics of the model is very regular and nonchaotic.

\section{Acknowledgments}

We are grateful to J. Ye, R. Liu, and J. Guo for helpful discussions. This work is supported by the Fujian Provincial Science Foundation under grant number 2016J05004.

\end{document}